# $k$-Colorability of $P_5$-free Graphs

C. Hoàng*   J. Sawada†   X. Shu‡

**Abstract**

A polynomial time algorithm that determines for a fixed integer $k$ whether or not a $P_5$-free graph can be $k$-colored is presented in this paper. If such a coloring exists, the algorithm will produce a valid $k$-coloring.

## 1 Introduction

Graph coloring is among the most important and applicable graph problems. The *k-colorability problem* is the question of whether or not the vertices of a graph can be colored with one of $k$ colors so that no two adjacent vertices are assigned the same color. In general, the $k$-colorability problem is NP-complete [13]. Even for planar graphs with no vertex degree exceeding 4, the problem is NP-complete [6]. However, for other classes of graphs, like perfect graphs [10], the problem is polynomial-time solvable. For the following special class of perfect graphs, there are efficient polynomial time algorithms for finding optimal colorings: chordal graphs [7], weakly chordal graphs [11], and comparability graphs [5]. For more information on perfect graphs, see [1], [3], and [9].

Another interesting class of graphs are those that are $P_t$-free, that is, graphs with no chordless paths $v_1, v_2, \ldots, v_t$ of length $t-1$. If $t = 3$ or $t = 4$, then there exists efficient algorithms to answer the $k$-colorability question (see [3]). However, for $t = 5, 6$, or 7, the complexity of the problem is unknown; for $t \geq 8$ the problem is NP-complete [16]. To handle the unknown cases, a natural first step is to consider what happens if the value of $k$ is fixed. Taking this parameterization into account, a snapshot of the known complexities for the $k$-colorability problem of $P_t$-free graphs is given in Table 1 (under columns 5 and 6, $\alpha$ is the exponent given by matrix multiplication that currently satisfies $2 < \alpha < 2.376$ [4]). From this chart we can see that there is a polynomial algorithm for the 3-colorability of $P_6$-free graphs [15].

In this paper we focus on $P_5$-free graphs. Notice that when $k = 3$, the colorability question for $P_5$-free graphs can be answered in polynomial time (see [16]). For $k = 4$ partial results were obtained in [12] when a $P_5$-free graph has a dominating $K_4$. The main result of this paper is the development of a polynomial time algorithm that will determine whether or not a $P_5$-free graph is $k$-colorable for a fixed integer $k$. If such a coloring exists, then the algorithm will yield a valid $k$-coloring. We note that if $k$ is part of the input, then the problem is NP-complete [14]. We also note the algorithm in [8] that colours a $(P_5, \overline{P_5})$-free graph $G$ with $\chi(G)^2$ colours, where $\chi(G)$ is the chromatic number of $G$.

*Physics and Computer Science, Wilfred Laurier University, Canada. Research supported by NSERC. email: choang@wlu.ca
†Computing and Information Science, University of Guelph, Canada. Research supported by NSERC. email: sawada@cis.uoguelph.ca
‡Computing and Information Science, University of Guelph, Canada. email: xshu@uoguelph.ca

| $k\backslash t$ | 3 | 4 | 5 | 6 | 7 | 8 | ... | 12 | ... |
|---|---|---|---|---|---|---|---|---|---|
| 3 | $O(m)$ | $O(m)$ | $O(n^\alpha)$ | $O(mn^\alpha)$ | ? | ? | ? | ? | ... |
| 4 | $O(m)$ | $O(m)$ | ?? | ? | ? | ? | ? | $NP_c$ | ... |
| 5 | $O(m)$ | $O(m)$ | ?? | ? | ? | $NP_c$ | $NP_c$ | $NP_c$ | ... |
| 6 | $O(m)$ | $O(m)$ | ?? | ? | ? | $NP_c$ | $NP_c$ | $NP_c$ | ... |
| 7 | $O(m)$ | $O(m)$ | ?? | ? | ? | $NP_c$ | $NP_c$ | $NP_c$ | ... |
| ... | ... | ... | ... | ... | ... | ... | ... | ... | ... |

Table 1: Known complexities for $k$-colorability of $P_t$-free graphs

The remainder of the paper is presented as follows. In Section 2 we present relevant definitions, concepts, and notations. Then in Section 3, we present our recursive polynomial time algorithm that answers the $k$-colorability question for $P_5$-free graphs.

## 2 Background and Definitions

In this section we provide the necessary background and definitions used in the rest of the paper. For starters, we assume that $G = (V, E)$ is a simple undirected graph where $|V| = n$ and $|E| = m$. If $A$ is a subset of $V$, then we let $G(A)$ denote the subgraph of $G$ induced by $A$.

DEFINITION 1 *A set of vertices A is said to* dominate *another set B, if every vertex in B is adjacent to at least one vertex in A.*

The following structural result about $P_5$-free graphs is from Bacsó and Tuza [2]:

THEOREM 1 *Every connected $P_5$-free graph has either a dominating clique or a dominating $P_3$.*

DEFINITION 2 *Given a graph G, an integer k and for each vertex v, a list $l(v)$ of k colors, the $k$-list coloring problem asks whether or not there is a coloring of the vertices of G such that each vertex receives a color from its list.*

DEFINITION 3 *The* restricted $k$-list coloring problem *is the $k$-list coloring problem in which the lists $l(v)$ of colors are subsets of $\{1, 2, \ldots, k\}$.*

Our general approach is to take an instance of a specific coloring problem $\Phi$ for a given graph and replace it with a polynomial number of instances $\phi_1, \phi_2, \phi_3, \ldots$ such that the answer to $\Phi$ is "yes" if and only if there is some instance $\phi_k$ that also answers "yes".

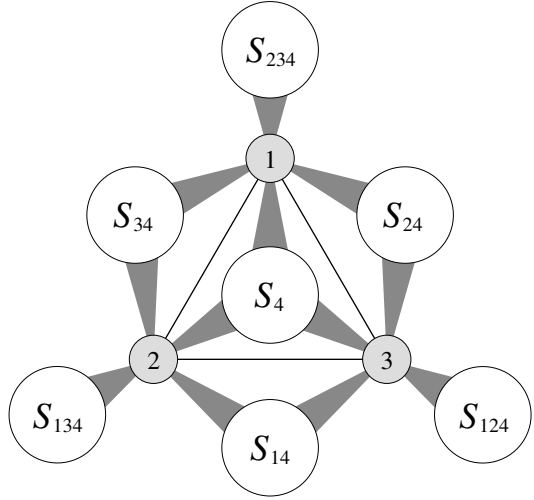

Figure 1: The fixed sets in a $P_5$-free graph with a dominating $K_3$ where $k = 4$.

For example, consider a graph with a dominating vertex $u$ where each vertex has color list $\{1, 2, 3, 4, 5\}$ This listing corresponds to our initial instance $\Phi$. Now, by considering different ways to color $u$, the following four instances will be equivalent to $\Phi$:

- $\phi_1$: $u = 1$ and the remaining vertices have color lists $\{2, 3, 4, 5\}$,
- $\phi_2$: $u = 2$ and the remaining vertices have color lists $\{1, 3, 4, 5\}$,
- $\phi_3$: $u = 3$ and the remaining vertices have color lists $\{1, 2, 4, 5\}$,
- $\phi_4$: $u = \{4, 5\}$ and the remaining vertices have color lists $\{1, 2, 3, 4, 5\}$.

In general, if we recursively apply such an approach we would end up with an exponential number of equivalent coloring instances to $\Phi$.

## 3 The Algorithm

Let $G$ be a connected $P_5$-free graph. This section describes a polynomial time algorithm that decides whether or not $G$ is $k$-colorable. Our strategy is as follows. First, we find a dominating set $D$ of $G$ which is a clique with at most $k$ vertices or a $P_3$. There are only a finite number of ways to colour the vertices of $D$ with $k$ colours. For each of these colourings of $D$, we recursively check if it can be extended to a colouring of $G$. Each of these subproblems can be expressed by a restricted list colouring problem. We now describe the algorithm in detail.

The algorithm is outlined in 3 steps. Step 2 requires some extra structural analysis and is presented in more detail in the following subsection.

1. Identify and color a maximal dominating clique or a $P_3$ if no such clique exists (Theorem 1). This partitions the vertices into **fixed sets** indexed by available colors. For example, if a $P_5$-free graph has a dominating $K_3$ (and no dominating $K_4$) colored with $\{1, 2, 3\}$ and $k = 4$, then the fixed sets would be given by: $S_{124}, S_{134}, S_{234}, S_{14}, S_{24}, S_{34}$. For an illustration, see Figure 1. Note that all the vertices in $S_{124}$ are adjacent to the vertex colored 3 and thus have color lists $\{1, 2, 4\}$. This gives rise to our original restricted list-coloring instance $\Phi$. Although the illustration in Figure 1 does not show it, it is possible for there to be edges between any two fixed sets.

2. Two vertices are *dependent* if there is an edge between them and the intersection of their color lists is non-empty. In this step, we remove all dependencies between each pair of fixed sets. This process, detailed in the following subsection, will create a polynomial number of coloring instances $\{\phi_1, \phi_2, \phi_3, \ldots\}$ equivalent to $\Phi$.

3. For each instance $\phi_i$ from Step 2 the dependencies between each pair of fixed sets has been removed which means that the vertices within each fixed set can be colored independently. Thus, for each instance $\phi_i$ we recursively see if each fixed set can be colored with the corresponding restricted color lists (the base case is when the color lists are a single color). If *one* such instance provides a valid $k$-coloring then return the coloring. Otherwise, the graph is not $k$-colorable.

As mentioned, the difficult part is reducing the dependencies between each pair of fixed sets (Step 2).

## 3.1 Removing the Dependencies Between Two Fixed Sets

Let $S_{list}$ denote a fixed set of vertices with color list given by $list$. We partition each such fixed set into **dynamic sets** $P_i$ that each represents a unique subset of the colors in $list$. For example: $S_{123} = P_{123} \cup P_{12} \cup P_{13} \cup P_{23} \cup P_1 \cup P_2 \cup P_3$. Initially, $S_{123} = P_{123}$ and the remaining sets in the partition are empty. However, as we start removing dependencies, these sets will dynamically change. For example, if a vertex $u$ is initially in $P_{123}$ and one of its neighbors gets colored 2, then $u$ will be removed from $P_{123}$ and added to $P_{13}$.

Recall that our goal is to remove the dependencies between two fixed sets $S_p$ and $S_q$. To do this, we remove the dependencies between each pair $(P, Q)$ where $P$ is a dynamic subset of $S_p$ and $Q$ is a dynamic subset of $S_q$. Let $col(P)$ and $col(Q)$ denote the color lists for the vertices in $P$ and $Q$ respectively. By visiting these pairs in order from largest to smallest with respect to $|col(P)|$ and then $|col(Q)|$, we ensure that we only need to consider each pair once. Applying this approach, the crux of the reduction process is to remove the dependencies between a pair $(P, Q)$ by creating at most a polynomial number of equivalent colorings.

Now, observe that there exists a vertex $v$ from the dominating set found in Step 1 of the algorithm that dominates every vertex in one set, but is not adjacent to any vertex in the other. This is because $P$ and $Q$ are subsets of different fixed sets. WLOG assume that $v$ dominates $Q$. Now, consider the (connected) components of $G(P)$ and $G(Q)$. If a component $Z$ in $G(P)$ is not adjacent to any vertex in $Q$ then the vertices in $Z$ have no dependencies with $Q$. The same applies for such components in $Q$. Since these components have no dependencies, we focus on the induced subgraph $H = G(P \cup Q \cup \{v\})$ with these components removed. This graph is illustrated in Figure 2 where the small rectangles represent the components in $G(P)$ and $G(Q)$ respectively. It is easy to observe that $H$ is connected (if not, then there are components $H_1, H_2$ of $H$, each of which contains a vertex in $P$ and a vertex in $Q$; it follows there are edges $(a, b)$ of $H_1$ and $(c, d)$ of $H_2$ such that $a, b, v, d, c$ induce a $P_5$).

THEOREM 2 *Let $H$ be a connected $P_5$-free graph partitioned into three sets $P$, $Q$ and $\{v\}$ where $v$ is adjacent to every vertex in $Q$ but not adjacent to any vertex in $P$. Then there exists at most one component in $G(P)$*

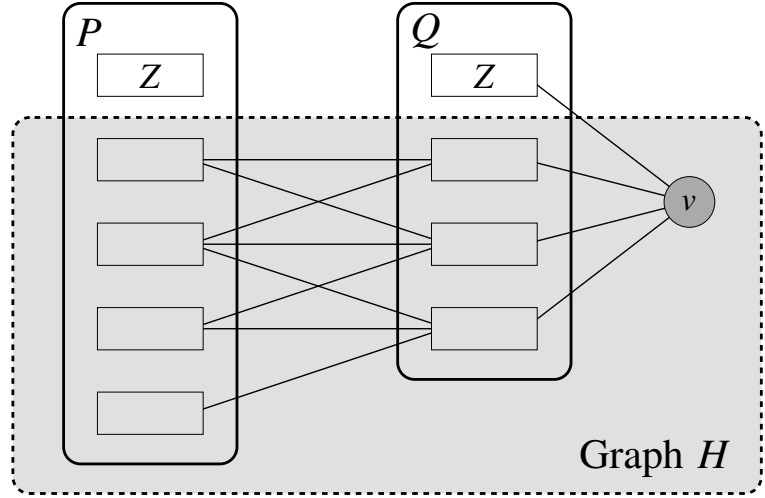

Figure 2: Illustration of the graph $H$ from two dynamic sets

*that contains two vertices $a$ and $b$ such that $a$ is adjacent to some component $Y_1 \in G(Q)$ but not adjacent to another component $Y_2 \in G(Q)$ while $b$ is adjacent to $Y_2$ but not $Y_1$.*

PROOF: The proof is by contradiction. Suppose that there are two unique components $X_1, X_2 \in G(P)$ with $a, b \in X_1$ and $c, d \in X_2$ and components $Y_1 \neq Y_2$ and $Y_3 \neq Y_4$ from $G(Q)$ such that:

- $a$ is adjacent to $Y_1$ but not adjacent to $Y_2$,
- $b$ is adjacent to $Y_2$ but not adjacent to $Y_1$,
- $c$ is adjacent to $Y_3$ but not adjacent to $Y_4$,
- $d$ is adjacent to $Y_4$ but not adjacent to $Y_3$.

Let $y_i$ denote an arbitrary vertex from the component $Y_i$. Since $H$ is $P_5$-free, there must be edges $(a, b)$ and $(c, d)$, otherwise $a, y_1, v, y_2, b$ and $c, y_3, v, y_4, d$ would be $P_5$s. An illustration of these vertices and components is given in Figure 3 - the solid lines.

Now, if $Y_2 = Y_3$, then there exists a $P_5 = a, b, y_2, c, d$. Thus, $Y_2$ and $Y_3$ must be unique components, and $Y_1, Y_4$ must be different as well for the same reason. Similarly $Y_2 \neq Y_4$. Now since $b, y_2, v, y_3, c$ cannot be a $P_5$, either $b$ is adjacent to $Y_3$ or $c$ must be adjacent to $Y_2$. WLOG, suppose the latter. Now $a, b, y_2, v, y_4$ implies that either $a$ or $b$ is adjacent to $Y_4$. If the latter, then $a, b, y_4, d, c$ would be a $P_5$ which implies that $a$ must be adjacent to $Y_4$ anyway. Thus, we end up with a $P_5 = a, y_4, v, y_2, c$ which is a contradiction to the graph being $P_5$-free. □

From Theorem 2, there is at most one component $X$ in $G(P)$ that contains two vertices $a$ and $b$ such that $a$ is adjacent to some component $Y_1 \in G(Q)$ but not adjacent to another component $Y_2 \in G(Q)$ while $b$ is adjacent to $Y_2$ but not $Y_1$. If such a component exists, then we can remove the vertices in $X$ from $P$ by applying the following general method for removing a component $C$ from a dynamic set $D$.

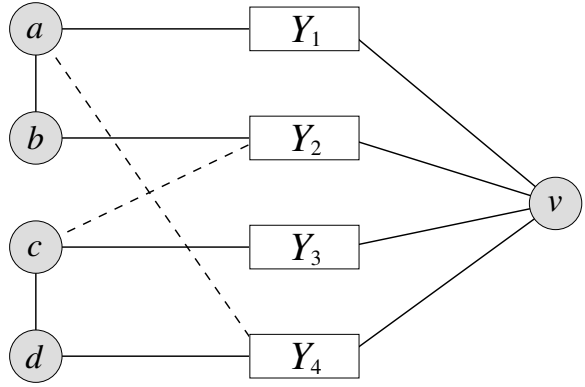

Figure 3: Illustration for proof of Theorem 2

### Remove Component

Since $C$ is $P_5$-free, it has a dominating clique or $P_3$ (Theorem 1). If this dominating set $D$ can be colored with the list $col(D)$, we consider all such colorings (otherwise we report there is no valid coloring for the given instance). For each case the coloring will remove all vertices in the component from $D$ to other dynamic sets represented by smaller subsets of available colors. Observe that since $k$ is fixed, the number of such colorings is constant.

If there are still dependencies between $P$ and $Q$, then we make the following claim (observing that the graph $H$ dynamically changes as $P$ and $Q$ change):

CLAIM 1 *There exists a vertex $x \in P$ that is adjacent to all components in $H(Q)$. Moreover, $x$ dominates all components of $H(Q)$ except at most one.*

PROOF: Let $x \in P$ be adjacent to a maximal number of components in $H(Q)$. If is not adjacent to all components, then there must exist another vertex $x' \in P$ and components $Y_1, Y_2 \in Q$ such that $x$ is adjacent to $Y_1$ but not $Y_2$ and $x'$ is adjacent to $Y_2$ but not $Y_1$. This implies that there is a $P_5 = x, y_1, v, y_2, x'$ where $y_1 \in Y_1$ and $y_2 \in Y_2$ unless $x$ and $x'$ are adjacent. However by Theorem 2, they cannot belong to the same component in $H(P)$ since such a component would already have been removed - a contradiction.

Now, suppose that there are two components $Y_1$ and $Y_2$ in $H(Q)$ that $x$ does *not* dominate. Then there exists edges $(y_1, y'_1) \in Y_1$ and $(y_2, y'_2) \in Y_2$ such that $x$ is adjacent to $y_1$ and $y_2$, but not $y'_1$ nor $y'_2$. This however, implies the $P_5 = y'_1, y_1, x, y_2, y'_2$ - a contradiction. □

Now we identify such an $x$ outlined in this claim and create equivalent new coloring instances by assigning $x$ with each color from $col(P) \cap col(Q)$ and then with the list $col(P) - col(Q)$. If $x$ is assigned a color from $col(P) \cap col(Q)$, then all but at most one component will be removed from $H(Q)$. If one component remains, then we can remove it from $Q$ by applying Remove Component. In the latter case, where $x$ is assigned the color list $col(P) - col(Q)$, $x$ will be removed from $P$. If there are still dependencies between $P$ and $Q$, we repeat this step by finding another vertex $x$. In the worst case we have to repeat this step at most $|P|$ times. Therefore, the process for removing the dependencies between two dynamic sets creates at most $O(n)$ new equivalent coloring instances.

Since there are a constant number of pairs of dynamic sets for each pair of fixed sets, and since there are constant number of pairs of fixed sets, this proves the following theorem:

THEOREM 3 *The restricted k-list coloring problem for $P_5$-free graphs, for a fixed integer k, can be solved in polynomial time.*

COROLLARY 1 *Determining whether or not a $P_5$-free graph can be colored with k-colors, for a fixed integer k, can be decided in polynomial time.*

We will now analyze the algorithm in more detail to provide a rough estimate of its complexity.

CLAIM 2 *Removing dependencies between two fixed sets produces $O(n^{2^{2(k-1)}})$ subproblems.*

*Proof.* As mentioned earlier, removing dependencies between two dynamic sets produces $O(n)$ subproblems. Each fixed set contains at most $O(2^{k-1})$ dynamic sets. Thus, between two fixed sets, there are $O(2^{2(k-1)})$ pairs of dynamic sets to consider. The claim now follows. □

There are $O(2^k)$ fixed sets. So there are $O(2^{2k})$ pairs of fixed sets. Thus, removing dependencies between all pairs of fixed sets produces $O(n^{2^{2(k-1)} 2^{2k}}) = O(n^{2^{4k-2}})$ subproblems, in each of these subproblems at most $k-1$ colors are used.

Let $T(k)$ be the number of subproblems produced by the algorithm with $k$ being the number of available colors. Then $T(k) = n^{2^{4k-2}} T(k-1)$ with $T(1) = 1$. An easy proof by induction shows $T(k) \leq (n^{2^{4k-2}})^k = n^{k \, 2^{4k-2}}$. Thus, the complexity of the algorithm in Theorem 3 is $O(n^{k \, 2^{4k-2}})$.